# Observations of the first confirmed superoutburst of SDSS J080434.20+510349.2 in 2006 March

Jeremy Shears, Geir Klingenberg & Pierre de Ponthière

**Abstract**

During 2006 March the first confirmed superoutburst of the dwarf nova SDSS J080434.20+510349.2 was observed using unfiltered CCD photometry. Time-series photometry revealed superhumps with a period of 0.0597 +/- 0.0011 d and an amplitude of 0.2 magnitude, thereby independently establishing its UGSU classification. Following the decline from a peak magnitude of 13.1, at least two rebrightening events were observed. Evidence is presented which is consistent with the star being a member of the UGWZ sub-class.

**Introduction**

Dwarf novae are a type of cataclysmic variable star in which a cool main sequence secondary star loses mass to a white dwarf primary. Material from the secondary falls through the inner Lagrangian point and, because it carries substantial angular momentum, does not settle on the primary immediately but forms an accretion disc. From time-to-time, as material builds up in the disc, thermal instability drives the disc into a hotter, brighter state causing an outburst in which the star brightens by several magnitudes. Dwarf novae of the SU UMa family (UGSU) occasionally exhibit superoutbursts which last several times longer than normal outbursts and may be up to a magnitude brighter. During a superoutburst the light curve of a UGSU star is characterized by superhumps. These are modulations which are a few percent longer than the orbital period and are thought to be caused by precession of the accretion disc [1]. Some researchers recognise a sub-class of UGSU stars which are more highly evolved systems; this sub-class is usually referred to as UGWZ after its prototype WZ Sge. UGWZ stars have long intervals between outbursts, typically decades, and exceptionally large outburst amplitudes, usually exceeding 6 magnitudes. For a full account of dwarf novae, and specifically UGSU and UGWZ stars, the reader is directed to reference 1.

**SDSS J080434.20+510349.2 and the Sloan Digital Sky Survey**

The Sloan Digital Sky Survey ("SDSS" [2,3]) is in the process of mapping a quarter of the sky, determining the positions and brightness of millions of objects. It is also measuring the distance to more than a million galaxies, which will allow researchers to determine the distribution of galaxies in space. The survey uses a dedicated 2.5 m telescope on Apache Point, New Mexico, USA, equipped with a 120 mega-pixel camera and a pair of spectrographs that can obtain spectra of hundreds of objects simultaneously. The SDSS completed it's first phase of operations in June 2005 having surveyed over 8,000 square degrees during which it detected 200 million objects and conducted spectroscopy on more than 675,000 galaxies, 90,000 quasars and 185,000 stars, some of which are suspected of being new dwarf novae. The second phase in now in progress and is due to be completed in June 2008.

Having identified so many objects, there is considerable follow up work underway to characterise them and conduct further investigations. One specific object, SDSS J080434.20+510349.2 (hereafter "SDSS0804"), was identified as a dwarf nova from studies of its spectrum by Paula Szkody and her team [4]. Photometry conducted at quiescence revealed some unusual behaviour: a double-humped orbital light curve, with a peak-to-peak amplitude of 0.05 mag and a rapid brightening by ~0.5 mag. The brightening was accompanied by an increase in amplitude of the orbital humps. They speculated that the brightening could be due to an increase in mass transfer or some sort of precession of the accretion disc.

The star is located in Lynx at 08h 04 min 34.2 s +51 deg 03 min 49.2 sec (J2000) and is approximately 17$^{th}$ magnitude in quiescence.

**Outburst light curve**

The outburst of SDSS0804 discussed in this paper was first detected by Elena Pavlenko at the Crimean Astronomical Observation on 2006 March 4 at approximately 19h UT using the 2.6 m telescope [5]. She reported that the star was around magnitude 13, compared with 17.5 when she had observed it in quiescence during 2006 January. Following the dissemination of the outburst announcement on various variable star distribution lists, the authors first observed the star at 13.1C on the evening (UT) of March 5 (Figure 1). Thus the outburst amplitude was at least 4.4 magnitudes.

Figure 2 shows the lightcurve of the outburst based on the authors' data plus data from the American Association of Variable Star Observers (AAVSO) International Database [6]. There was a rapid decline during the first two days of observation to 15.0C, suggesting that the outburst was already well advanced when discovered by Pavlenko. The decline was followed by at least 2 rebrightening events and a subsequent decline to 16.8C, still above quiescence, 81 days after the outburst was first detected. Pavlenko and her co-workers actually reported eleven rebrightening events based on a more complete data set in a detailed paper on this star [7]. Unfortunately relatively few observers have reported observations to the AAVSO database, resulting in an incomplete observational record which explains why several of the rebrightening events have been missed.

**Detection of superhumps**

The authors conducted time-series photometry of SDSS0804 on March 5 using the instrumentation shown in Table 1 and according to the observation log in Table 2. The resulting light curve (Figure 3) reveals the presence of 6 superhumps, thereby confirming this object to be a member of the UGSU class of dwarf nova. The peak-to-peak superhump amplitude was ~ 0.2 magnitudes. This is an independent confirmation of the presence of superhumps reported by Pavlenko *et al.* on March 4 [7].

Further time-series data on March 6 showed that the star was fading rapidly at ~1.1 mag/d (Figure 4). Superhumps were still present, but they were less obvious and with diminished amplitude. In fact, there is some evidence from Figure 3 that the final superhump in the previous night's run was already exhibiting less coherence that the preceding ones. We also note that there are rapid variations of a few tenths of a magnitude, especially towards the end of the observation. This appears to be flickering which is commonly observed in outbursts of dwarf novae. A similar loss of coherent superhumps, accompanied by intense flickering, was reported during the outburst of the dwarf nova 1RXS J053234.9+624755 Cam [8]. Poor weather prevented further time-series observations by the authors.

**Superhump and orbital periods**

The time-series data from March 5 (Figure 3) were analysed for periodicities using the CLEANest algorithm in Peranso [9]. Given the apparent loss of integrity during the final superhump mentioned in the previous section, we decided to eliminate observations made after JD 2453800.597 from our analysis. Figure 5 shows the resulting power spectrum. The strongest signal arises from the superhumps and reveals a superhump frequency of 16.741 +/- 0.295 c/d and a corresponding superhump period $P_{sh}$ = 0.0597 +/- 0.0011 d. The uncertainty is calculated using the Schwarzenberg-Czerny method [10]. This value is consistent with the one reported by Pavlenko *et al.* of 0.059713 +/- 0.000007 d [7].

Folding the combined time-series data on $P_{sh}$ gives the phase diagram in Figure 6. Careful inspection of the phase diagram reveals a small hump at a phase of ~ 0.8. The hump can also just be made out in the descending phase of the superhumps in Figure 3. A similar hump was noticed by Pavlenko *et al.* during the outburst [7] and by Szkody *et al.* during quiescence [4]. This may well be an orbital hump associated with the bright spot formed at in the region where material from the secondary hits the edge of the accretion disc. Such bright spots have been observed in many dwarf novae including U Gem [1], WZ Sge [11] and Z Cha [12].

Analysis of the data from March 6 (Figure 4) using Peranso failed to establish a significant period within the data. As stated above, the superhumps were poorly defined during this observation and flickering was present.

Szkody *et al.* proposed an orbital period, $P_{orb}$, for SDSS0804 of 0.0590 +/- 0.0020 d from both spectroscopic and photometric measurements [4]. Pavlenko *et al.* measured $P_{orb}$ as 0.0586 +/- 0.0008 d based on photometry during quiescence [7]. We note that the period quoted by each researcher lies within the error quoted by the other researcher, i.e. given the errors involved, both results are consistent with each other.

We also note that removing $P_{sh}$ from the power spectrum in Figure 5 leaves a weak signal at 0.0295 +/- 0.007 d, which is more or less half the value of Szkody *et al.*'s $P_{orb}$ [4]. It is possible that this relates to the orbital hump at a phase of ~ 0.8 described above. However, given the errors involved in the measurements of $P_{sh}$ and $P_{orb}$, there is also the possibility that this signal is simply a harmonic of $P_{sh}$. Which of these two possibilities is correct could only be concluded from longer data runs.

As stated in the Introduction, the superhump period of UGSU systems is usually slightly longer than the orbital period, the difference being defined as the "period excess", $\varepsilon$, where $\varepsilon = (P_{sh} - P_{orb}) / P_{orb}$. If we take $P_{sh}$ for SDSS0804 as 0.0597 d, we calculate the following period excess values:

      based on Szkody *et al.*'s value of $P_{orb}$ [4]      $\varepsilon$ = 0.0119
      based on Pavlenko *et al.*'s value of $P_{orb}$ [7]    $\varepsilon$ = 0.0188

The period excess of a number of UGSU systems is related to the orbital period: the shorter the orbital period, the smaller the period excess [13]. A star with one of the shortest orbital periods is WZ Sge, the prototype of the UGWZ sub-class of dwarf novae.

This has $P_{orb}$ = 0.0567 and ε = 0.0080. Szkody *et al.* [4] have suggested that SDSS0804 is a member of the UGWZ sub-class based on spectroscopic evidence and the very short orbital period. UGWZ systems are more highly evolved than the majority of UGSU systems and the optical spectrum is dominated by a cool white dwarf, with no secondary star visible.

A particular feature of UGWZ stars is that they have a very long outburst period of several years to decades: for example, WZ Sge itself has an outburst every 33 years or so. The factor that determines the outburst period appears to be the mass-transfer rate. A typical UGSU system has a mass-transfer rate of about $5 \times 10^{12}$ kg/s (~$10^{-10}$ solar mass/year), whereas for UGWZ systems it is perhaps only $10^{12}$ kg/s [1]. It then takes many years to accumulate enough material to trigger a superoutburst. In the case of SDSS0804 there is evidence from the inspection of plate archives of only one outburst, other than the present one, which occurred in 1979 when the star was found at approximately magnitude 12.5 [7].

So is SDSS0804 a member of the UGWZ class? Taking into account the spectroscopic evidence for a highly evolved system [4], the short orbital period [4, 7], the low period excess (present study) and the infrequent outbursts [7], this appears to be almost certain.

We also note that several UGWZ systems exhibit rebrightening events, or "echo" outbursts, following the initial outburst and which are very similar to those we have reported above in SDSS0804. For example the UGWZ star EG Cnc exhibited six echo outbursts during its 1996 outburst [14] and WZ Sge itself exhibited 12 such echoes [11].

How does the outburst amplitude of SDSS0804 compare with UGWZ stars? As mentioned in the Introduction, UGWZ stars typically exhibit large outburst amplitudes, usually exceeding 6 magnitudes, compared to 4-6 magnitudes for UGSU systems. For example, WZ Sge, HV Vir, EG Cnc and AL Com all have amplitudes in the range of ~ 6 to 8 magnitudes [11, 15, 14, 16]. The 4.4 magnitude amplitude for SDSS0804 mentioned above is more modest, but it should be emphasised that the actual amplitude may well be greater. There is uncertainty over both the maximum brightness (13.1), since it may have been brighter before it was first detected, and the quiescence brightness. Szkody *et al.* [4] point out that it appears that SDSS0804 varies between 17.6 and 18.3 during quiescence. It we accept the latter as the true minimum, the amplitude then increases to *at least* 5.2 magnitudes.

**SDSS0804 and the Recurrent Objects Programme**

The BAA Variable Star Section's Recurrent Objects Programme (ROP) was set up specifically to encourage the monitoring of poorly studied eruptive stars of various types, especially those with long outburst periods [17, 18]. SDSS0804 was added to the ROP by the co-ordinator Gary Poyner in 2006, following the current outburst. Although it appears that the previous outburst of SDSS0804 was back in 1979, it is possible that others have been missed and thus it may not mean that we have to wait a further 27 years for another one. Hence further monitoring of this star is strongly encouraged by both visual observers and those equipped with CCDs. To assist observers who may wish to take up this challenge, a preliminary chart for SDSS0804 appears in Figure 7.

**Acknowledgements**

The authors gratefully acknowledge the use of observations from the AAVSO International Database contributed by observers worldwide and we thank Elizabeth Waagen of the AAVSO for making the verified data available to us for this study. We are indebted to Dr. Elena Pavlenko for helpful discussions and for providing a pre-print of reference 7, to Dr. Boris Gaensicke for helpful comments on the draft of this paper, as well as for his encouragement. We thank Gary Poyner, co-ordinator of the BAA Variable Star Section's Recurrent Objects Programme, for permission to reproduce the preliminary chart of SDSS0804 and for preparing it for publication. We acknowledge the use of SIMBAD, operated through the Centre de Données Astronomiques (Strasbourg). Finally, we thank the referees for constructive comments.

**Addresses:**
JS: "Pemberton", School Lane, Bunbury, Tarporley, Cheshire, CW6 9NR, UK [bunburyobservatory@hotmail.com]
GK : Bossmo Observatory, Mo i Rana, Norway [geir.klingenberg@gmail.com]
PdP: Lesve-Profondeville, Belgium [pierredeponthiere@gmail.com]

| Observer | Instrumentation | Exposure time (s) |
|---|---|---|
| GK | 0.2-m Newtonian reflector + SBIG ST-7 CCD | 90 |
| PdP | 0.2-m SCT + SBIG ST7XEI CCD | 30 |
| JS | 0.1-m fluorite refractor + SXV-M7 CCD | 60 |

**Table 1: Instrumentation used**

| Date (2006) | Start time (JD) | Duration (hours) | No. of images | Mean mag (unfiltered) | Observer |
|---|---|---|---|---|---|
| March 5 | 2453800.299 | 7.2 | 714 | 13.16 | PdP |
| March 5 | 2453800.306 | 9.2 | 262 | 13.22 | GK |
| March 5 | 2453800.324 | 2.0 | 112 | 13.18 | JS |
| March 6 | 2453801.325 | 7.9 | 547 | 13.50 | PdP |

**Table 2: Log of time-series observations**
In all cases the photometric comparison star was GSC 3414:1011 (mag 11.38)

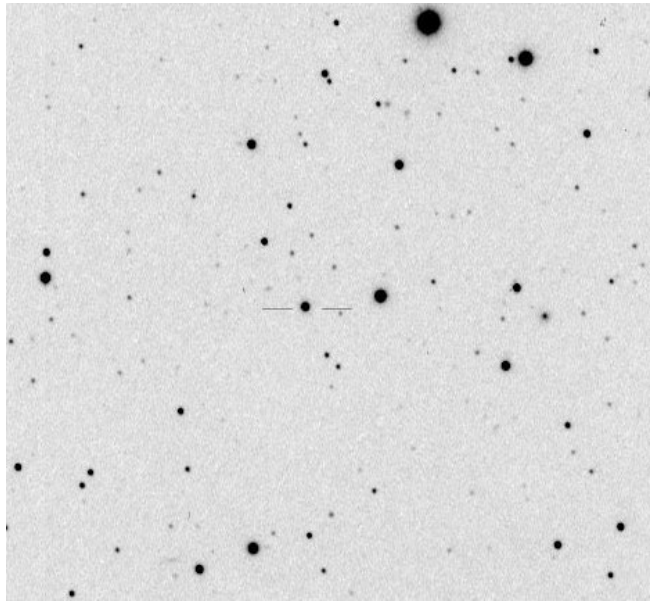

**Figure 1: SDSS0804 on 2006 March 5, 19.47UT**
Field 14.5 x 14.5 arcmin, N at bottom, E to right

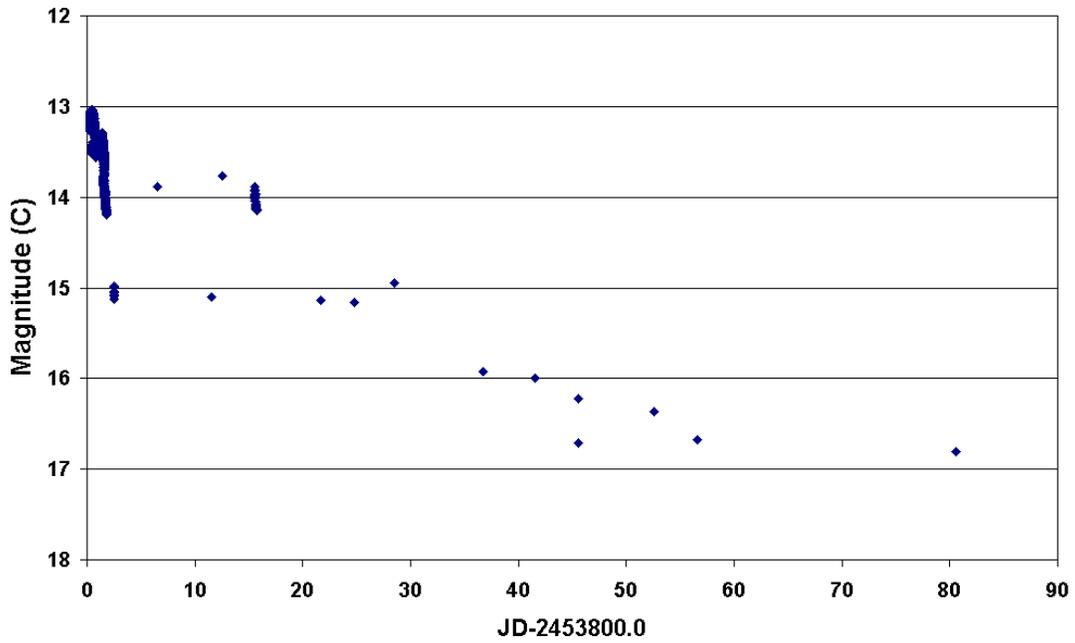

**Figure 2: Light curve of the outburst**
Unfiltered CCD photometry from the authors and the AAVSO International Database

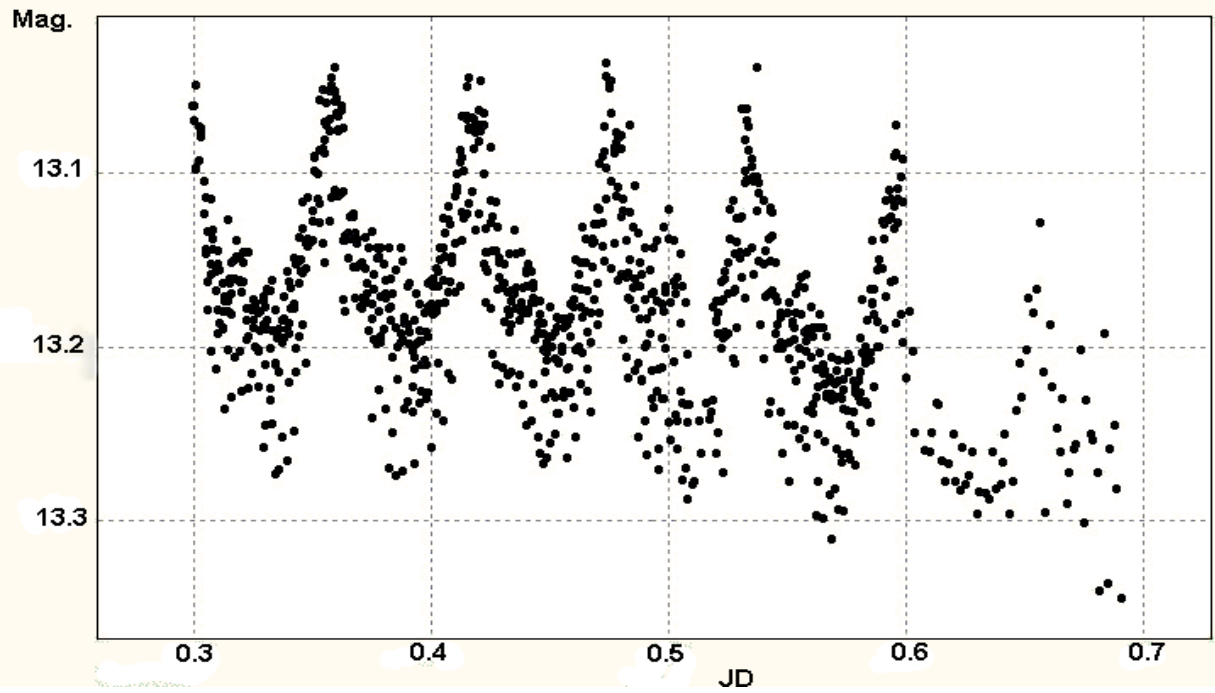

**Figure 3: Detection of superhumps on 2006 March 5
(JD-2453800.0)**

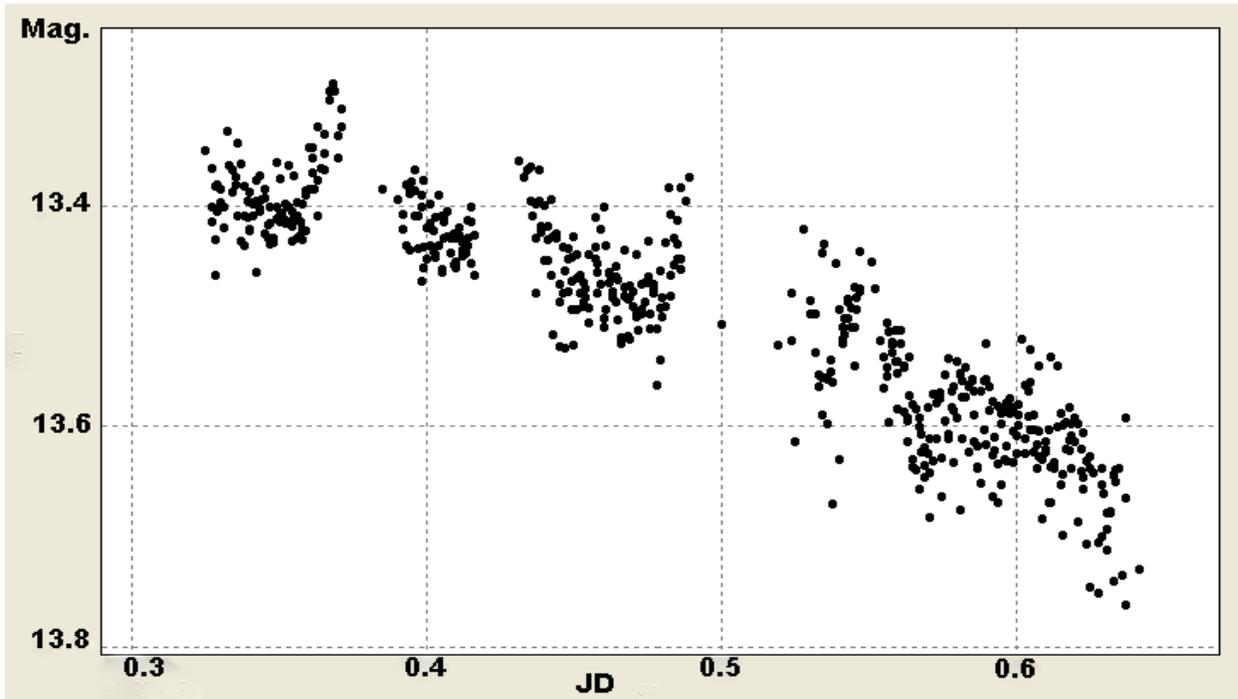

**Figure 4: Time resolved photometry on 2006 March 6
(JD-2453801.0)**
Gaps in the data are due to cloud cover

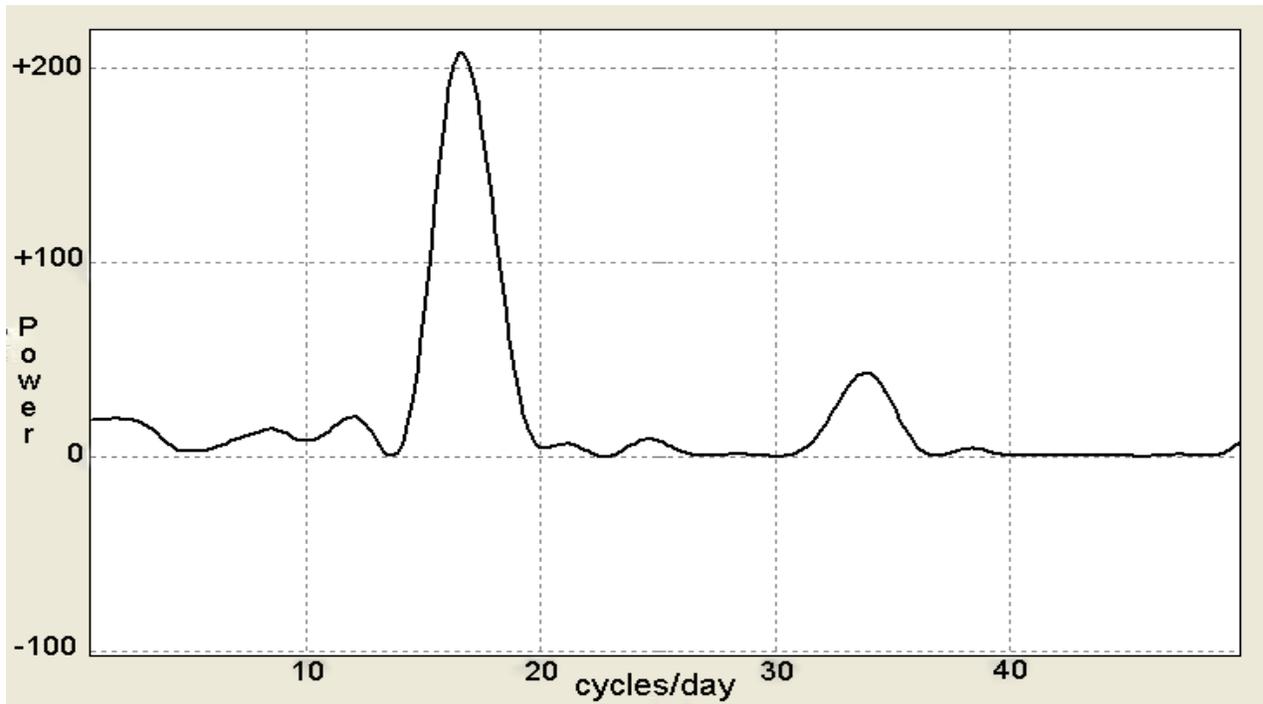

**Figure 5: Power spectrum of the data from 2006 March 5**
(Data after JD 2453800.597 were excluded from our analysis)

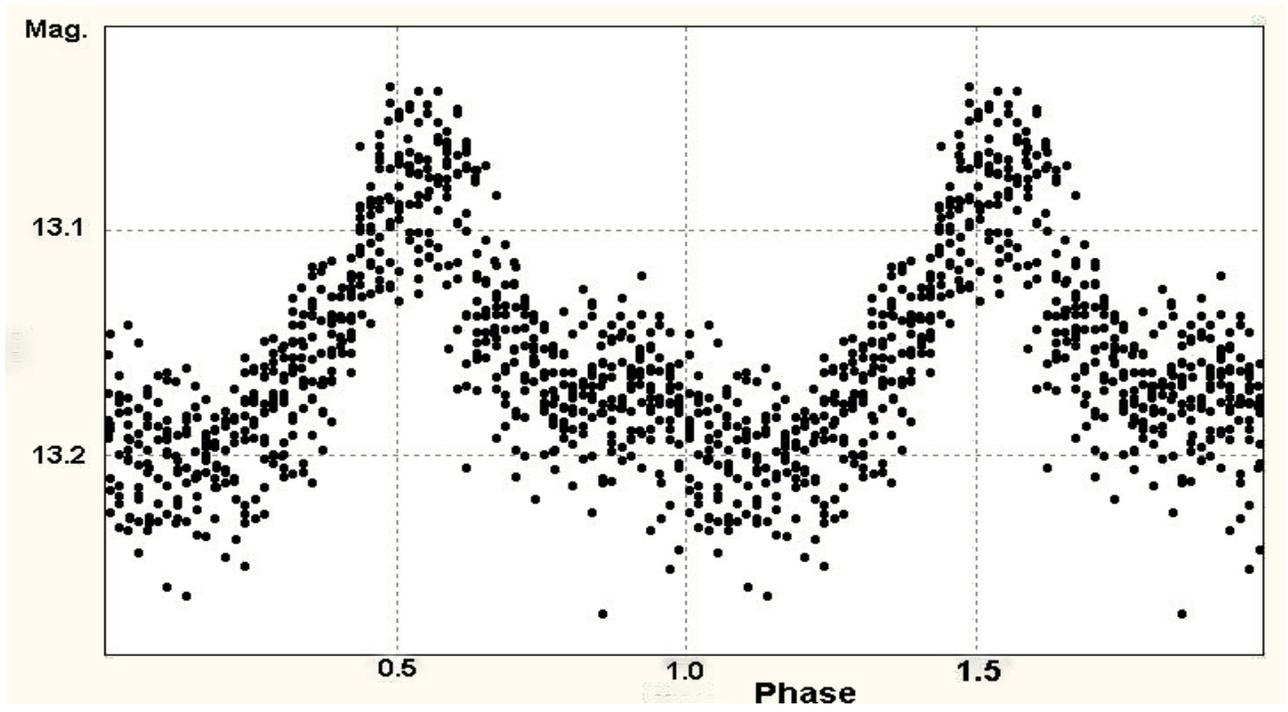

**Figure 6: Phase diagram of time-series data from 2006 March 5 folded on P$_{sh}$**

# SDSSJ080434.20+510349.2

RA 08 04 34.2   Dec +51 03 49.2
12.0-18.3V   Type: UGWZ?

20' Field

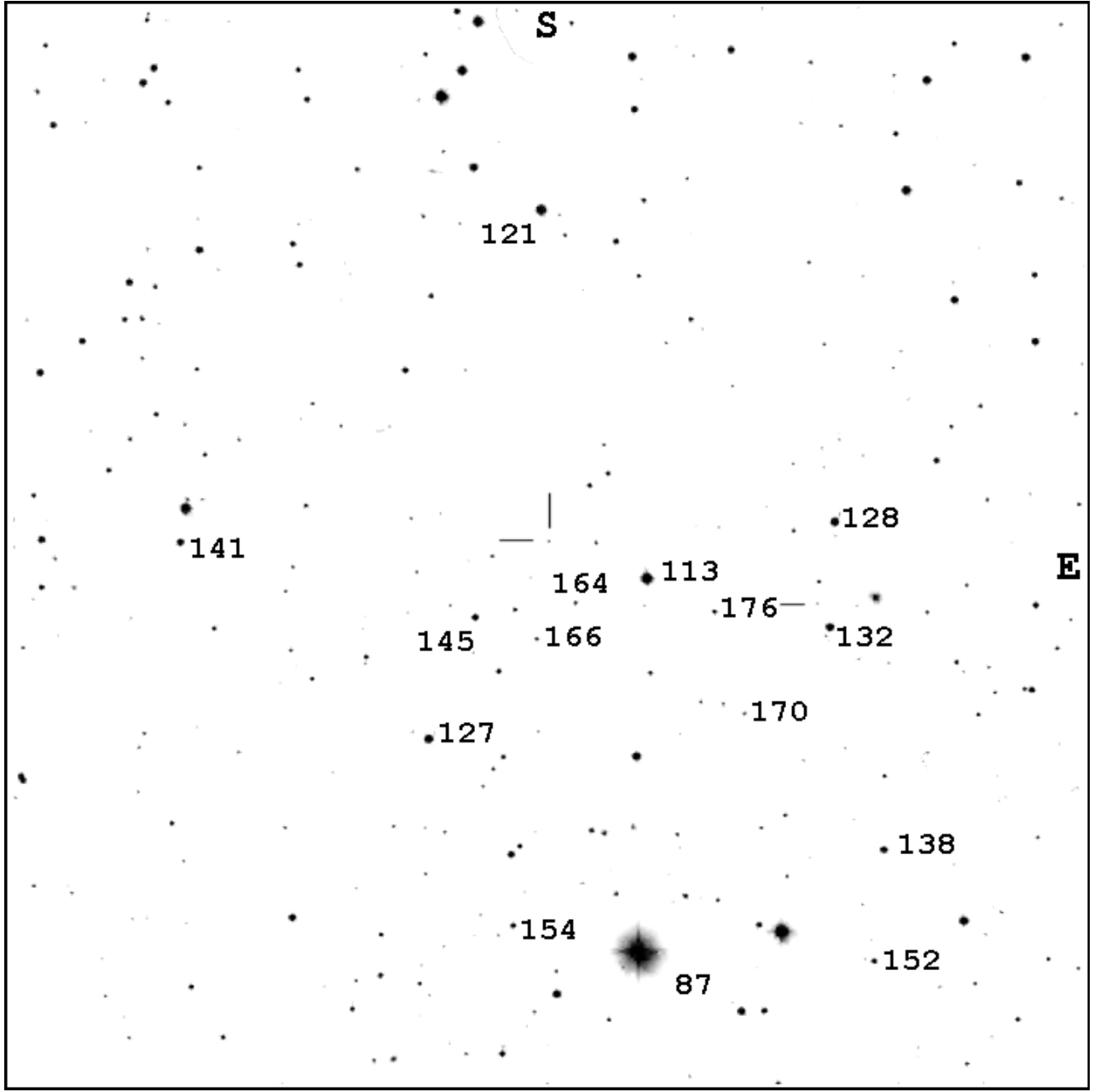

Seq.  USNO A2.0

**Figure 7: Chart for SDSS0804**